\documentclass[fleqn,usenatbib]{mnras}

\usepackage{newtxtext,newtxmath}

\usepackage[T1]{fontenc}

\DeclareRobustCommand{\VAN}[3]{#2}
\let\VANthebibliography\thebibliography
\def\thebibliography{\DeclareRobustCommand{\VAN}[3]{##3}\VANthebibliography}

\usepackage{graphicx}	
\usepackage{amsmath}	
\usepackage{xcolor}
\usepackage[normalem]{ulem}  

\newcommand{\WI}[2]{#1_{\mathrm{#2}}}

\newcommand{\re}{\WI{R}{e}}
\newcommand{\rp}{\WI{R}{p}}
\newcommand{\rL}{\WI{r}{L_1}}

\newcommand{\Mtot}{\WI{M}{tot}}
\newcommand{\orb}{\WI{\Omega}{orb}}

\title[Dynamics of direct impact accretion]{Dynamics of direct impact accretion in degenerate binary systems}

\author[N. Kramarev and A. Yudin]{
Nikita Kramarev,$^{1,2}$\thanks{E-mail: kramarev-nikita@mail.ru}
Andrey Yudin$^{1}$
\\
$^{1}$ National Research Center Kurchatov Institute,
    pl. Kurchatova 1, Moscow, 123182, Russia\\
$^{2}$ Lomonosov Moscow State University, Sternberg Astronomical Institute,
    Universitetsky pr. 13, Moscow, 119234, Russia
}

\date{Accepted XXX. Received YYY; in original form ZZZ}

\pubyear{2023}

\begin{document}
\label{firstpage}
\pagerange{\pageref{firstpage}--\pageref{lastpage}}
\maketitle

\begin{abstract}
We consider the gas dynamics in an accreting binary system of degenerate stars within the framework of the Newtonian approximation. In such a system, the accretion stream can impact the surface of a white dwarf (WD) or neutron star (NS) as a result of the very compact orbit. This causes a loss of angular momentum from the orbit and spin-up of the accretor. We construct approximation for the specific angular momentum of the accreting matter which goes to spin up the accretor and approximations for some other parameters of the system. It is shown that the obtained approximation of the specific momentum is qualitatively different from the widely used Keplerian formula. It should affect the boundary between scenarios of immediate tidal disruption and slow mass loss of the donor in WD-WD and NS-NS binaries, as well as the time of stable mass transfer in the stripping scenario.
\end{abstract}




\section{Introduction}
Binary systems of degenerate stars with mass transfer are of great interest for astrophysicists as progenitors of Type Ia supernovae \citep[WD-WD systems][]{Iben1984ApJS,Webbink1984ApJ} and short gamma-ray bursts \citep[NS-NS, or NS-BH systems][]{Lattimer1974ApJ}. If the radius of the accretor is comparable to its Roche radius, then mass transfer stream can hit the accretor directly \citep[e.g.][]{LubowShu1975}. This scenario is realised during the coalescence of a double degenerate binary system with high mass asymmetry, for example, during accretion in a WD-WD system at the final stages of its evolution. In this case, orbital angular momentum of the system is effectively transferred to the accretor that shifts the mass ratio limit for a dynamically stable mass transfer \citep{Nelemans2001a,Nelemans2001b,Marsh2002,Marsh2004}. Accounting for spin-up can also change the duration of the mass transfer phase that influences the details of the accretor surface detonation \citep{Guillochon2010,Dan2011,Dan2012,Dan2015} and the subsequent Type Ia supernova explosion \citep[e.g.][]{Kashyap2018}.

The transfer of orbital angular momentum to rotational momentum of the accretor during the direct impact-phase should also be taken into account when considering NS-NS coalescence in the stripping model for short gamma-ray bursts \citep{ClarkEardley1977,Blinnikov1984,Eichler1989,Blinnikov1990}. The calculations of \citet{Blinnikov2022} have shown that this effect causes a decrease in the duration of the stable mass transfer $\WI{t}{str}$ (or the stripping time) --- the most important dynamical parameter of the NS stripping model. As discussed in \citet{Blinnikov2021}, $\WI{t}{str}$ can be attributed to the time delay between the peak of the gravitational-wave signal GW170817 and gamma-ray burst GRB 170817A detection \citep{Abbott2017a,Abbott2017b}.

In many analytical calculations of WD-WD system evolution during direct impact accretion the Keplerian formula is used for the specific angular momentum \citep[e.g.][]{Marsh2002,Marsh2004,2007ApJ.Gokhale}. The other approach involves the numerical solution of restricted three-body problem \citep{WarnerPeters1972,Flannery1975,Sepinsky2010ApJ} to obtain the desired value of angular momentum. Of course, these calculations of the mass transfer process is much more time-consuming.

We will show that the results of both aforementioned approaches differ not only quantitatively but also qualitatively \citep[see also ][]{Sepinsky2014ApJ}. The main goal of our work is to propose a simple approximation for the specific angular momentum of the accreting matter to speed up the calculations. Besides we obtain few auxiliary approximations of some important parameters of the system.

This paper is structured as follows: in Section \ref{2} we derive equations for the specific angular momentum of the accreting matter which goes to spin up WD or NS during direct impact accretion. We also present analytical solutions to the restricted three-body problem in two limiting cases. On the basis of these solutions, in Section \ref{3} we obtain the desired approximation of the specific momentum. Some auxiliary approximations are also obtained. We conclude the paper in Section \ref{4} with a comparison of the obtained approximation for the angular momentum with the Keplerian formula.

\section{Formalism}
\label{2}
\subsection{Basic Assumptions}

We consider a close binary system of NS-NS (or WD-WD) with masses $M_1$ (accretor) and $M_2$ (donor). The binary is assumed to be in a circular Keplerian orbit with the distance between stars $a$ and orbital rotation frequency of the system $\Omega_{\mathrm{orb}} {=} \sqrt{G\WI{M}{tot}/{a^3}}$, where $\Mtot=M_1+M_2$. It is justified by the population synthesis calculations \citep{Kowalska2011} and the gravitational-wave observations for NS-NS systems \citep{Lenon2020}, allowing to neglect of orbital eccentricity in most cases.

We also assume the mass of each star is distributed spherically symmetrically. This is fully justified for the donor by the fact that the stable mass transfer requires sufficient asymmetry of the binary system component masses \citep{Marsh2004,Blinnikov2022}. In fact, the low-mass component consists of a very dense core, which contains nearly the entire mass, and an extended light envelope \citep{Koester1990,Haensel_book}. Moreover, as shown by \citet{Yudin2020} for NS-NS binaries, even if there is an initial angular momentum, the low-mass component loses it quickly enough during accretion. Therefore, we consider that the donor corotates with the orbit.

However, the sphericity can be violated for the extremely rotating massive component (see Appendix \ref{Appendix0}). The contribution of the general relativity (GR) effects can also be significant near the surface of the accretor (see Appendix \ref{Appendix}). We will take both non-sphericity and GR effects into account in the subsequent work.

\subsection{Accretion spin-up of the massive component}

During stable mass transfer from the donor surface to the accretor, the orbital angular momentum of the binary system is also transferring to the rotational momentum of NS (or WD) $J_1$. Taking into account the condition of mass conservation, these can be written as
\begin{equation}
\dot{J}_1=-\dot{M}_2 \mathfrak{j}(q,r_1)a^2\orb,\label{J1_dot}
\end{equation}
where $\mathfrak{j}$ is the specific angular momentum of matter in orbital units. Here we define $r_1=R_1/a$ as the dimensionless stopping radius,
where $R_1$ is the equatorial radius of the accretor. In the next subsections we show that within the framework of the Newtonian approximation, $\mathfrak{j}(q,r)$ can be represented as a function of two parameters: the ratio of the donor mass to the total mass of the system, $q=M_2/\WI{M}{tot}$\footnote{Note that our definition differs from the more common relation $q'=M_2/M_1$.}, and the dimensionless distance to the centre of mass of the accretor, $r=R/a$. Due to the properties of the mass-radius dependence of degenerate stars \citep[e.g.][]{ShapiroTeukolsky_book}, during accretion in NS-NS and WD-WD systems, the accretor is always a massive component and the donor is a low-mass one, so in our case $q\leq 0.5$.

\begin{figure}
    \includegraphics[width=\columnwidth]{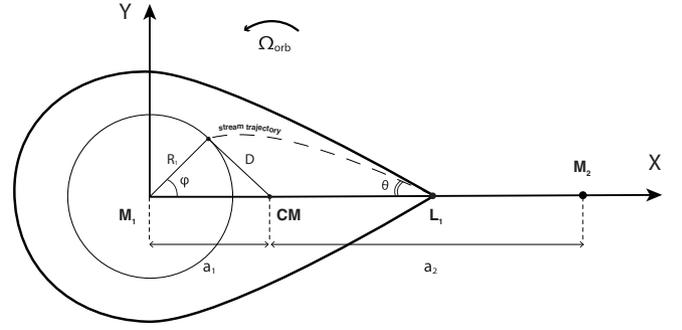} 
	\caption{Schematic diagram of the rotating Cartesian coordinate system $(X,Y,Z{=}0)$ whose centre coincides with the position of the massive component. The symbol $\mathrm{CM}$ marks the centre of mass of the system, and the dashed curve is the stream trajectory starting from the inner Lagrange point, L1, at an angle $\theta$ and hitting the surface of the accretor at a point with polar coordinates $(R_1,\varphi)$.} \label{accretion_plot}
\end{figure}

Let us consider the kinematics of direct impact accretion of matter on the surface of the accretor leading to the spin-up of the latter. First, we deduce the relationship between the angular momentum $\mathfrak{j}$ in the inertial frame (IF) of reference and its value $j$ in the synchronously rotating frame (RF) from purely kinematic considerations. The IF is connected to the centre of mass of the binary system (see Fig.~\ref{accretion_plot}).
Let the speed of the stream in the RF when it hits the surface of a massive component be given by the vector $\vec{V}$ while the speed of the accretor is zero. In the IF they will be added by the velocities $[\vec{\Omega}_{\mathrm{orb}}\times \vec{D}]$ and $[\vec{\Omega}_{\mathrm{orb}}\times \vec{a}_1]$, respectively. The vectors $\vec{D}$ and $\vec{a}_1$ are drawn from the centre of mass of the system to the point where the stream hits the accretor and the centre of mass of the accretor, respectively. From the law of conservation of momentum we obtain:
\begin{equation}
M_1[\vec{\Omega}_{\mathrm{orb}}\times \vec{a}_1]+\triangle M_1\left(\vec{V}+[\vec{\Omega}_{\mathrm{orb}}\times \vec{D}]\right)=(M_1{+}\triangle M_1)\vec{V}_1,\label{impuls}
\end{equation}
where $M_1$ and $\triangle M_1$ are the masses of the massive component and the falling matter, respectively, and $\vec{V}_1$ is the velocity of NS (or WD) after the impact. The law of conservation of angular momentum gives us:
\begin{eqnarray}
M_1\big[\vec{a}_1\times[\vec{\Omega}_{\mathrm{orb}}\times \vec{a}_1]\big]+\vec{J}_{\mathrm{i}}+\triangle M_1\big[\vec{D}\times(\vec{V}+[\vec{\Omega}_{\mathrm{orb}}\times \vec{D}])\big]= \nonumber \\
=(M_1{+}\triangle M_1)[\vec{a}_1\times \vec{V}_1]+\vec{J}_{\mathrm{f}},\label{ugl_moment}
\end{eqnarray}
where $\vec{J}_{\mathrm{i,f}}$ is rotational momentum of the accretor $\vec{J}_1$ before and after the impact. Expressing $\vec{V}_1$ from (\ref{impuls}), we obtain
\begin{equation}
\triangle\vec{J}_1=\triangle M_1[\vec{R}_1\times \vec{V}]+\triangle M_1 \vec{\Omega}_{\mathrm{orb}}\big(D^2{-}(\vec{a}_1 \cdot\vec{D})\big).
\end{equation}
The first term on the right-hand side is the angular momentum of the matter with mass $\triangle M_1$ in the RF associated with the accretor. The second term represents the kinematic correction which can be easily expressed from geometric considerations (see Fig.~\ref{accretion_plot}). Then we can rewrite the spin-up equation in the form:
\begin{equation}
\dot{J}_1=\dot{M}_1\orb\Big(a^2 j + R_1(R_1{-}a_1\cos\varphi)\Big).\label{spinningNS}
\end{equation}
Comparing the formulae (\ref{J1_dot}) and (\ref{spinningNS}) and also taking into account $a_1=a q$, we get the desired relationship between the specific angular momentum of the accreting matter in different frames:
\begin{equation}
\mathfrak{j}(q,r_1)=j(q,r_1)+r_1\left(r_1{-}q \cos{\varphi}\right).\label{j_tot}
\end{equation}

Let us once again pay attention to the following circumstance: we obtained the expression for the specific angular momentum of the accreting matter from purely kinematic considerations. So it does not matter to us how the transferred moment is distributed in the accretor, whether it rotates differentially or in a rigid body - all these effects do not influence the evolution of the system. In the systems of interest to us, the rate of mass-transfer is $\dot{M}_1\sim \WI{M}{\odot}/$sec for NS-NS \citep{ClarkEardley1977} and $\dot{M}_1\sim 10^ {-3}\WI{M}{\odot}/$year for WD-WD \citep{Marsh2004} so the effects associated with the interaction of the magnetic field of the accretor with the gas stream are also insignificant \citep[e.g.][]{LipunovPostnov1984}. In our approach, the NS (or WD) spin-up is determined by the mass ratio $q$ and the dimensionless stopping radius $r_1$. To understand how $j$ and $\cos{\varphi}$ in expression (\ref{j_tot}) depend on $q$ and $r$, it is necessary to know the stream trajectory, i.e. to solve the restricted three-body problem.

\subsection{The motion of matter in the ballistic approach}

We now consider the dynamics of a stream of matter in the Roche lobe of the accretor in the ballistic approximation. This is justified by the supersonic nature of the flow, which makes it possible to neglect the effects of pressure. We also assume that matter leaves the Lagrange point, L1, with sufficiently small (or zero) velocity in the RF. The specific value and direction of the initial velocity vector does not influence the dynamics of the stream \citep[e.g.][]{Flannery1975}.

We choose a coordinate system as shown in Fig.~\ref{accretion_plot}. In the plane $(X,Y,Z{=}0)$ the radius vector $\vec{R}$ has components $(R\cos{\varphi},R\sin{\varphi})$ and the velocity $ \vec{V}=(V_x,V_y)=(\dot{R}\cos{\varphi}{-}R\Omega\sin{\varphi},\dot{R}\sin{\varphi}{+ }R\Omega\cos{\varphi})$ where we also introduce the notation $\Omega\equiv\dot{\varphi}$. 
The components of accelerations acting on a particle of matter are
\begin{align}
&\vec{a}_{M_1}=-\frac{G M_1}{R^2}(\cos{\varphi},\sin{\varphi}),\\
&\vec{a}_{M_2}=\frac{G M_2}{R_2^{3}}(a{-}R\cos{\varphi},{-}r\sin{\varphi}),\\
&\vec{a}_{\mathrm{R}}=\orb^2(r\cos{\phi}{-}a_1,r\sin{\phi}),\\
&\vec{a}_{\mathrm{C}}=2\orb(V_y,{-}V_x),
\end{align}
where the gravitational forces acting from $M_1$ and $M_2$, the centrifugal force, and the Coriolis force are taken into account. For brevity we also introduce the notation $R_2=\sqrt{R^2+a^2-2aR\cos{\varphi}}$. The equations of motion in polar coordinates can be written as \cite[see e.g.][]{1999MurrayDermott.book}:
\begin{align}
\ddot{R}-R\Omega^2&={-}\frac{G M_1}{R^2}+\frac{G M_2}{R_2^3}(a\cos{\varphi}{-}R) + \nonumber\\
&+(R{-}a_1\cos{\varphi})\orb^2+2 R \Omega \orb,\label{energy1}\\
2\dot{R}\Omega+R\dot{\Omega}&={-}\frac{GM_2}{R_2^3}a\sin{\varphi}+\orb^2 a_1\sin{\phi}-2 \dot{R} \orb.\label{moment1}
\end{align}

We now express all distances in units of $a$ and time in units of $t_0=1/\orb=\sqrt{a^3/GM}$. Then the equations (\ref{energy1}-\ref{moment1}) take the dimensionless form:
\begin{align}
\ddot{r}-r\omega^2&={-}\frac{1{-}q}{r^2}+\frac{q}{r_2^3}(\cos{\varphi}{-}r)+(r{-}q\cos{\varphi})+2\omega r,\label{energy2}\\
2\dot{r}\omega+r\dot{\omega}&={-}\frac{q}{r_2^3}\sin{\varphi}+q\sin{\varphi}-2\dot{r},\label{moment2}
\end{align}
where $r=R/a$, $\omega=\Omega t_0$, $r_2=\sqrt{1+r^2-2r\cos{\varphi}}$. Multiplying the equation (\ref{moment2}) by $r$ we bring it to the form:
\begin{equation}
\frac{d}{dt}\left(r^2(1{+}\omega)\right)=qr\sin{\varphi}\Big(1-\frac{1}{r_2^3}\Big).\label{moment3}
\end{equation}
This is obviously the momentum balance equation. By multiplying the equation (\ref{energy2}) by $\dot{r}$ and taking into account (\ref{moment3}) it can be integrated and brought to the form:
\begin{equation}
\frac{1}{2}\left[\dot{r}^2+(r\omega)^2\right]=\frac{1{-}q}{r}+\frac{q}{r_2}+\frac{r^2}{2}-qr\cos{\varphi}+\WI{E}{J},\label{energy3}
\end{equation}
where $\WI{E}{J}=\mathrm{const}$ is the Jacobi integral which is the only known integral of motion in the restricted circular three-body problem \citep[e.g.][]{Gurfil_book}.

We are not interested in the behavior of the quantities under consideration over time, so we exclude the time derivative according to $\frac{d}{dt}=\omega\frac{d}{d\phi}$. This allows us to finally write the system of equations in the form:
\begin{align}
\frac{\omega^2}{2}\Big[\Big(\frac{dr}{d\varphi}\Big)^2+r^2\Big]&=(1{-}q)\Big[\frac{1}{r}-\frac{1}{\rL}\Big]+q\Big[\frac{1}{r_2}{-}\frac{1}{1{-}\rL}\Big]+\nonumber\\
&+\frac{r^2{-}\rL^2}{2}+q(\rL{-}r\cos{\varphi}),\label{energy4}\\
\omega\frac{d}{d\varphi}\left(r^2(1{+}\omega)\right)&=qr\sin{\varphi}\Big(1-\frac{1}{r_2^3}\Big).\label{moment4}
\end{align}
To determine $\WI{E}{J}$, we take into account the initial condition that matter leaves the Lagrange point, L1, with zero velocity. The coordinate $\rL$ of the L1 point is determined from the equation:
\begin{equation}
\frac{1{-}q}{\rL^3}=1+q\frac{2{-}\rL}{(1{-}\rL)^2},\label{L1-r}
\end{equation}

We need to get the dependencies $r=r(\varphi)$ and $\omega=\omega(\varphi)$ with initial conditions $r(0)=\rL$ and $\omega(0)=0$ in order to further define $j=\omega r^{2}$ and $\cos{\varphi}$. So the equations (\ref{energy4}) and (\ref{moment4}) should be integrated up to the point $r_1=R_1/a$ where $R_1$ is the equatorial radius of the rotating accretor.

\subsection{The exact solutions}

The equations (\ref{energy4}-\ref{moment4}) have analytical solutions in the L1 region and around the point of minimal approach of the stream to $M_1$. This circumstance helps us to later construct approximations of $j$ and $\cos{\varphi}$ as functions of $q$ and $r$.

Let us first consider the motion of matter near the point L1 where $r \sim \rL$. For convenience, we place the origin of the rotating coordinate system $(x,y,z)$ at L1, so that the $x$-axis is directed to $M_1$ and the positive $y$-axis points out of the half-plane of the motion of the matter. The Roche potential in dimensionless coordinates up to quadratic terms is
\begin{equation}
\WI{\Phi}{R}=\frac{y^2}{2}f(q)-x^2\left[\frac{3}{2}+f(q)\right],
\end{equation}
where we also determine the function
\begin{equation}
f(q)=q\frac{\rL^2-3\rL+3}{(1-\rL)^3}.
\end{equation}
The acceleration of the particle consists of the acceleration in the Roche potential $\WI{\vec{a}}{R}=-\nabla\WI{\Phi}{R}$ and the Coriolis force, which looks like $\WI{\vec{a}}{K}=(2\dot{y},-2\dot{x})$ in component-wise notation with dimensionless variables.
The equations of motion can then be written as follows:
\begin{align}
\ddot{x}&=\big[3+2f(q)\big]x+2\dot{y},\\
\ddot{y}&=-yf(q)-2\dot{x}.
\end{align}
Since this is a system of linear differential equations, we are looking for a solution in the form of $x,y\sim e^{\alpha t}$. The system compatibility condition leads to the equation for $\alpha$:
\begin{equation}
\alpha^4-(f-1)\alpha^2-f(3+2f)=0,
\end{equation}
whose solution is
\begin{equation}
\alpha^2=\frac{1}{2}\left[f-1\pm\sqrt{(f+1)(9f+1)}\right].\label{alpha}
\end{equation}
We search for an infinitely growing solution that is a positive root of (\ref{alpha}) corresponding to the plus sign. Then, the asymptotic stream angle that the stream leaves the L1 region is $\tan\theta=\left|\frac{y(t)}{x(t)}\right|_{t\rightarrow\infty}\!\!=\frac{2\alpha}{f+\alpha^2}$. More compact is the expression for the cosine of the angle $\theta$:
\begin{equation}
\cos^2\theta=\frac{3f-1+\sqrt{(f+1)(9f+1)}}{6(f+1)}.\label{theta}
\end{equation}
So after leaving the point L1, the matter under the influence of the Coriolis force moves in a straight line at an angle $\theta$ to the line connecting the centres of the stars (see Fig.~\ref{accretion_plot}). This expression for the angle was first found in \citet{LubowShu1975}.

In polar coordinates centred at the point $M_1$, the trajectory of the gas stream in the L1 region has the form
\begin{equation}
r=\frac{\rL\sin\theta}{\sin(\theta{+}\varphi)}.\label{r_L1}
\end{equation}
Taking into account the dependence of $r\sim e^{\alpha t}$ it can be shown that $j$ can be explicitly expressed as a function of $r$:
\begin{equation}
j=\alpha\rL\sin\theta\left[\rL\cos{\theta}-\sqrt{\rL^2\cos^2\theta-\rL^2+r^2}\right].\label{j_L1}
\end{equation}

Now let us find the asymptotics of the trajectory in the region of the minimum approach of the stream and the massive component. We use two factors obtained from the analysis of the numerical solution of the system (\ref{energy4}-\ref{moment4}). First, we will assume that the angular momentum $j$ is approximately conserved here so that $\omega=\WI{\omega}{m}\left(\frac{\WI{r}{m}}{r} \right)^2$. The subscript $\mathrm{m}$ corresponds to the values at the point of minimum approach. On the right side (\ref{energy4}) we save only the first term and some constant which is determined from the conditions at the point of minimum approach. Such a choice is dictated by the expansion of the right-hand side (\ref{energy4}) in the small parameter $r/\rL$ for $r\sim\WI{r}{m}\ll\rL<1$. Then the equation (\ref{energy4}) reduces to
\begin{equation}
\WI{\omega}{m}^2\left(\frac{\WI{r}{m}}{r}\right)^4\left[r^2+\Big(\frac{dr}{d\varphi}\Big)^2\right]=2(1{-}q)\left(\frac{1}{r}-\frac{1}{\WI{r}{m}}\right)+\WI{\omega}{m}^2\WI{r}{m}^2.
\end{equation}
This equation can be simply integrated:
\begin{equation}
r=\frac{\WI{r}{m}}{1-\Lambda[1{-}\cos(\WI{\varphi}{m}{-}\varphi)]},\label{r_m_phi}
\end{equation}
where we introduce
\begin{equation}
\Lambda=1-\frac{1-q}{\WI{r}{m}^3\WI{\omega}{m}^2}.
\end{equation}
After all these assumptions, it is not surprising that we have obtained the usual equation of a conic section in polar coordinates. Near the massive component, all other forces except for its gravitation are small corrections and the motion occurs along the arc of an ellipse or hyperbola.

\section{Approximations}
\label{3}

It is time-consuming to solve the system of differential equations (\ref{energy4}-\ref{moment4}) at each time step in order to determine the angular momentum $\mathfrak{j}(q,r{=}r_1)$ transferred to the rotational momentum of a massive NS (or WD). To speed up the further calculation of the mass transfer process, we construct approximations of $j$ and $\cos{\varphi}$ included in the expression (\ref{j_tot}) for the momentum $\mathfrak{j}$ as functions of $q$ and $r$. At first, we consider auxiliary approximations of the radii $\rL(q)$, $\WI{r}{m}(q)$ and the angle $\WI{\varphi}{m}(q)$.

\subsection{Approximation of $\rL(q)$}

\begin{figure}
	\includegraphics[width=\columnwidth]{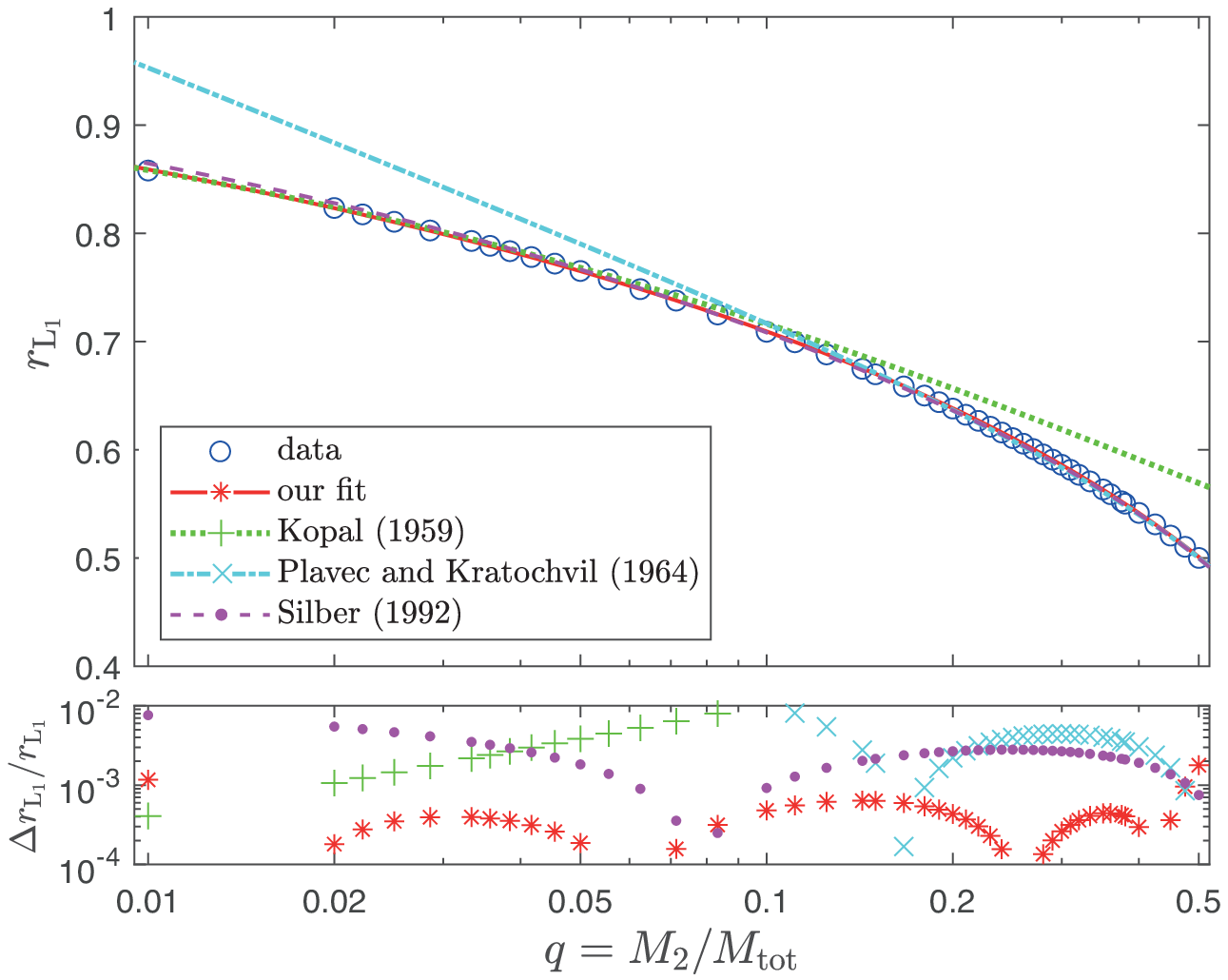}
	\caption{Upper panel: Comparison of various approximations of $\rL(q)$ where numerical data is shown in blue circles. Lower panel: Deviations between analytically fitted curves and computed data points ($\Delta \rL = |\mathrm{fit}-\mathrm{data}|$).} \label{Fig.rL_approx}
\end{figure}

The dimensionless distance between the inner Lagrangian point, L1, and centre of the accretor $\rL$ as a function of mass ratios $q$ is determined in (\ref{L1-r}). We propose the approximation
\begin{equation}
\rL=\left(a_1+a_2 q^{n_1}\right)^{n_2}, \; 0.01\leqslant q \leqslant 0.5 \label{rL_approx}
\end{equation}
where the coefficients are
\begin{center}
	\begin{tabular}{|c|c|}
		\hline
		$a_1$ & 1.334\\ 
		\hline
		$a_2$ & -1.29\\ 
		\hline
		$n_1$ & 1.462$\cdot 10^{-1}$\\ 
		\hline
		$n_2$ & 3.88$\cdot 10^{-1}$\\ 
		\hline
	\end{tabular}
\end{center}
The Fig.~\ref{Fig.rL_approx} shows numerical data of $\rL$ as a function of $q$ (blue circles) and various approximations taken from \citet{Kopal_book}, \citet{Plavec1964} and \cite{Silber1992}. Our curve fitting data, better than $0.2\%$, is shown in red. In a slightly narrower range, $ 0.02\leqslant q \leqslant 0.45$, its accuracy reaches $0.06\%$.

\subsection{Approximation of $\WI{r}{m}(q)$ and $\WI{\varphi}{m}(q)$}

\begin{figure}
	\includegraphics[width=\columnwidth]{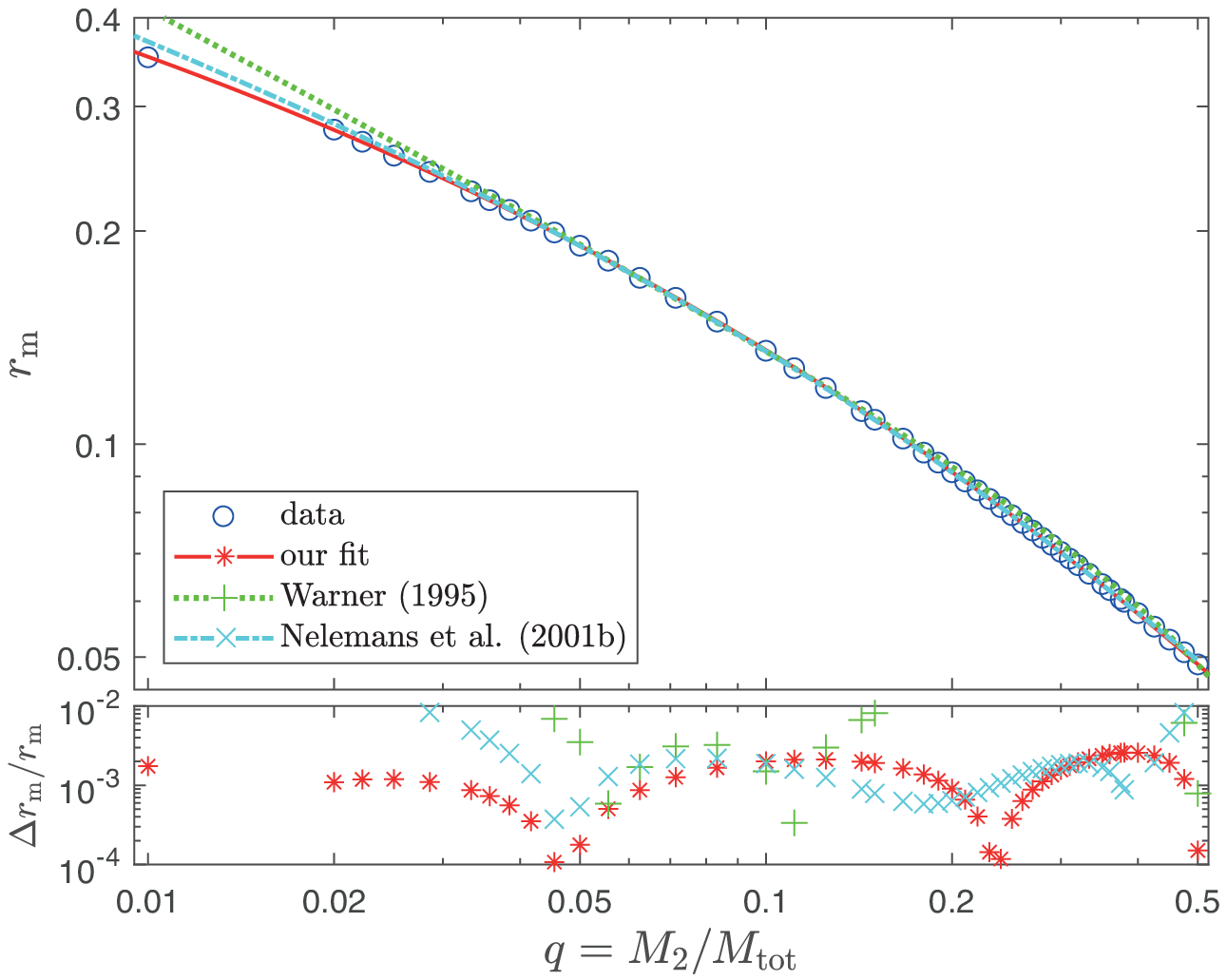}
	\caption{Upper panel: Comparison of various approximations of $\WI{r}{m}(q)$ where numerical data is shown in blue circles. Lower panel: Deviations between analytically fitted curves and computed data points ($\Delta \WI{r}{m} = |\mathrm{fit}-\mathrm{data}|$).} \label{Fig.rm_approx}
\end{figure}

For the minimum distance of approach of the stream to the massive component $\WI{r}{m}=\WI{R}{m}/a$, we propose the following approximation:
\begin{equation}
\WI{r}{m}=\exp\left(a_1+a_2 q^{n_1}\right), \; 0.01\leqslant q \leqslant 0.5 \label{rm_approx}
\end{equation}
where the coefficients are
\begin{center}
	\begin{tabular}{|c|c|}
		\hline
		$a_1$ & 3.951$\cdot 10^{-1}$\\ 
		\hline
		$a_2$ & -3.98\\ 
		\hline
		$n_1$ & 2.211$\cdot 10^{-1}$\\ 
		\hline
	\end{tabular}
\end{center}
Numerical data for $\WI{r}{m}$ are shown in Fig.~\ref{Fig.rm_approx} in blue circles. The approximations taken from \citet{Warner_book} and \citet{Nelemans2001b} are the green and cyan lines, respectively, and our approximation is the red line.

\begin{figure}
	\includegraphics[width=\columnwidth]{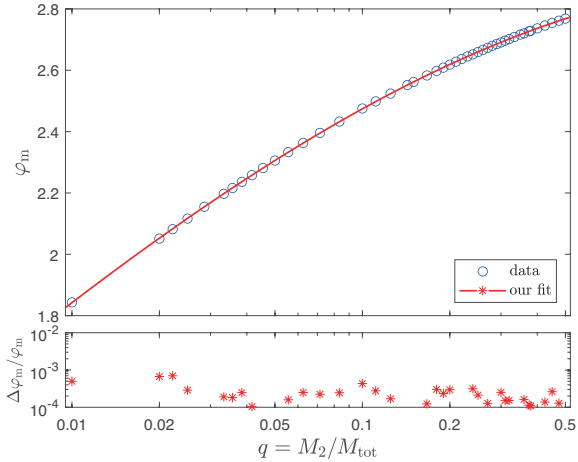}
	\caption{Upper panel: Comparison of numerical data with our approximation of $\WI{\varphi}{m}(q)$. Lower panel: Deviations between our fit and computed data points ($\Delta \WI{\varphi}{m} = |\mathrm{fit}-\mathrm{data}|$).} \label{Fig.phi_min}
\end{figure}

We also need an approximation of the angle $\WI{\varphi}{m}$ at $r=\WI{r}{m}$ as a function of $q$ that we chose in the following form (see also Fig.~\ref{Fig.phi_min})
\begin{equation}
\WI{\varphi}{m}=\sum\limits_{i=0}^3 a_{i}(\ln{q})^i, \; 0.01\leqslant q \leqslant 0.5\label{phi_min_approx}
\end{equation}
where the coefficients are
\begin{center}
	\begin{tabular}{|c|c|}
		\hline
		$a_0$ & 2.845\\ 
		\hline
		$a_1$ & 8.63$\cdot 10^{-2}$ \\ 
		\hline
		$a_2$ & -3.635$\cdot 10^{-2}$\\ 
		\hline
		$a_3$ & -1.695$\cdot 10^{-3}$\\
		\hline
	\end{tabular}
\end{center}
The general accuracy of the presented approximations is better than $0.5\%$.

\subsection{Approximation of $j(q,r)$}

\begin{figure}
	\includegraphics[width=\columnwidth]{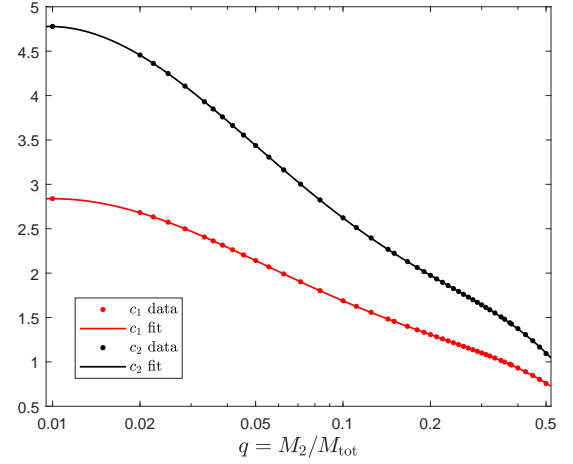}
	\caption{The data for $c_{1,2}(q)$ (dots) and approximations (lines).} \label{Fig_c12_approx}
\end{figure}

Here we find a convenient approximation for $j=\omega r^2$. Previously, we considered two limiting cases of the motion of the stream and showed that when approaching $\WI{r}{m}$, the specific angular momentum is approximately preserved, and in the L1 region it is described by formula (\ref{j_L1}). This suggests looking for an approximation in the form
\begin{equation}
j=c_1(q)-\sqrt{c_2(q)[r^2{-}\rL^2]+c_1^2(q)},\label{j_approx}
\end{equation}
where we take into account that $j=0$ at $r=\rL$ and use formula (\ref{rL_approx}) for $\rL=\rL(q)$. We select for the coefficients $c_{1,2}=c_{1,2}(q)$ polynomial in the $\ln{q}$ approximation:
\begin{equation}
c_k=\sum\limits_{i=0}^5 a_{k i}(\ln{q})^i.\label{c_12_coeff}
\end{equation}
Approximation coefficients (\ref{c_12_coeff}) are shown in the table below.
\begin{center}
	\begin{tabular}{|c|c|c|}
		\hline
		$i$ & $a_{1i}$ & $a_{2i}$\\
		\hline
		0 & -2.124$\cdot 10^{-1}$ & -4.995$\cdot 10^{-1}$\\ 
		\hline
		1 & -2.154 & -3.597 \\
		\hline
		2 & -1.425 & -2.455\\
		\hline
		3 & -5.561$\cdot 10^{-1}$& -9.669$\cdot 10^{-1}$\\
		\hline
		4 & -9.423$\cdot 10^{-2}$& -1.616$\cdot 10^{-1}$\\
		\hline
		5 & -5.518$\cdot 10^{-3}$& -9.196$\cdot 10^{-3}$\\
		\hline
	\end{tabular}
\end{center}
The data for $c_{1,2}$ in the range $0.01\leq q\leq 0.5$ (dots) and the approximation (\ref{c_12_coeff}) (lines) are shown in Fig.~\ref{Fig_c12_approx}.

\begin{figure}
	\includegraphics[width=\columnwidth]{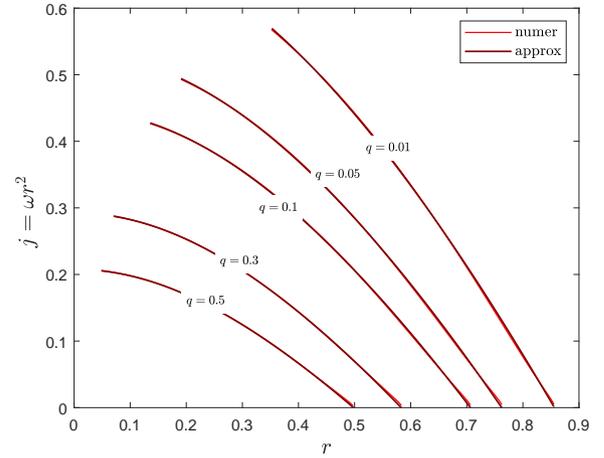}
	\caption{Numerical data for $j=j(r)$ and their approximation for several values of $q$.} \label{Fig_j_r}
\end{figure}

Comparison of the results of the numerical calculation $j=j(r)$ and approximation (\ref{j_approx}) is shown in Fig.~\ref{Fig_j_r}. The thin red lines ending at the points $r=\WI{r}{m}(q)$ show the calculation for several values of $q$, and the thick brown lines show the approximation. The accuracy of our approximation is much better than $1\%$ in almost the entire range of $\WI{r}{m}\leqslant r \leqslant \rL$ and $0.01\leqslant q \leqslant 0.5$.

\subsection{Approximation of $\cos{\varphi}(q,r)$}

Now we find a suitable approximation for $\cos{\varphi}$. Taking into account the limiting cases (\ref{r_L1}) for $\varphi \approx 0$ and (\ref{r_m_phi}) for $\varphi \rightarrow \WI{\varphi}{m}$, we look for an approximation in the form
\begin{equation}
\frac{1}{r}=(a_0+a_1\sin\varphi+a_2\cos\varphi)^{1/n}.\label{r_phi}
\end{equation}
Comparison with formula (\ref{r_L1}) with $r=\rL$ up to terms of order $O(\varphi)$ gives two equations
\begin{align}
&a_0+a_2=\frac{1}{\rL^n},\\
&a_1=n\frac{\cot\theta}{\rL^n},
\end{align}
where $\theta$ is considered in (\ref{theta}). Two more relations follow from the conditions $r(\WI{\varphi}{m})=\WI{r}{m}$ and $(dr/d\varphi)_{\varphi{=}\WI{\varphi }{m}}=0$. This leads to the equations
\begin{align}
&\frac{1}{\WI{r}{m}^n}=a_0+a_1\sin\WI{\varphi}{m}+a_2\cos\WI{\varphi}{m},\\
&a_1\cos\WI{\varphi}{m}=a_2\sin\WI{\varphi}{m}.
\end{align}

Finally, it all comes down to a single equation for $n$:
\begin{equation}
\left(\frac{\rL}{\WI{r}{m}}\right)^n-1=n\cot\theta\frac{1{-}\cos\WI{\varphi}{m}}{\sin\WI{\varphi}{m}}.\label{n_eqn}
\end{equation}
We solve it and provide an approximation with an accuracy better than $0.3\%$
\begin{equation}
n=\frac{a_1+q^{n_1}}{a_2+q^{n_2}}, \; 0.01\leqslant q \leqslant 0.5\label{n_approx}
\end{equation}
where the coefficients are
\begin{center}
	\begin{tabular}{|c|c|}
		\hline
		$a_1$ & 1.535$ \cdot 10^{-1}$ \\ 
		\hline
		$a_2$ & -6.8003$\cdot 10^{-2}$ \\ 
		\hline
		$n_1$ & 3.312$\cdot 10^{-1}$\\ 
		\hline
		$n_2$ & 3.166$\cdot 10^{-1}$\\ 
		\hline
	\end{tabular}
\end{center}

The remaining parameters can be found by explicit formulae
\begin{align}
a_0&=\frac{1{-}n\cot\theta\cot\WI{\varphi}{m}}{\rL^n},\\
a_1&=n\frac{\cot\theta}{\rL^n},\\
a_2&=n\frac{\cot\theta\cot\WI{\varphi}{m}}{\rL^n}. \label{a2_approx}
\end{align}
It can be seen that the terms with sine and cosine in (\ref{r_phi}) are grouped to form $\cos(\varphi{-}\WI{\varphi}{m})$, so the solution of this equation with respect to $\varphi$ at known $r$ can be written as
\begin{equation}
\varphi=\WI{\varphi}{m}-\bigg|\arccos{\Bigg(\frac{1/r^n{-}a_0}{\sqrt{a_1^2{+}a_2^2}}\bigg)}\Bigg|. \label{phi_approx1}
\end{equation}
Substituting the formulas for the parameters $a_0,$ $a_1$, and $a_2$ we finally get
\begin{equation}
\varphi=\WI{\varphi}{m}-\bigg|\arccos{\Bigg[\cos\WI{\varphi}{m}+\frac{\sin\WI{\varphi}{m}}{n\cot\theta}\Bigg(\bigg(\frac{\rL}{r}\bigg)^n\!\!{-}1\Bigg)\Bigg]}\Bigg|. \label{phi_approx2}
\end{equation}

\begin{figure}
	\includegraphics[width=\columnwidth]{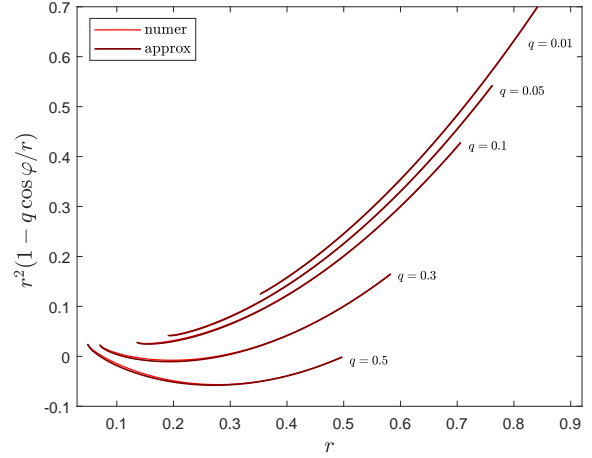}
	\caption{Numerical data for the kinematic correction and its approximation for several values of $q$.} \label{Fig.cos_phi}
\end{figure}

It is important to note that the approximation obtained for $\cos\varphi$, strictly speaking, is not an approximation because its coefficients are obtained not from fitting the numerical solution, but from analytical solutions for two limiting cases. Fig.~\ref{Fig.cos_phi} shows the kinematic correction from the general formula $(\ref{j_tot})$ for the specific momentum of matter $\mathfrak{j}$. For the function $\cos{\varphi}$ from $r$, we take the numerical solution and the solution obtained by formulae (\ref{n_approx})-(\ref{phi_approx2}). The relatively low accuracy of our approximation near $\cos\varphi \approx 0$ is offset both by the smallness of the cosine of the angle itself, and by the factor $q$ in the general formula for $\mathfrak{j}$, so the relative accuracy of the entire kinematic correction is better than 1\%.

Thus, we managed to find relatively simple two-parametric approximations for $j$ and $\cos{\varphi}$ in the ranges $0.01\leq q\leq 0.5$ and $\WI{r}{m}\leq r\leq \rL $. This eliminates the need to solve the system (\ref{energy4}-\ref{moment4}) each time in order to find the specific angular momentum $\mathfrak{j}(q,r{=}r_1)$ transferred to the accretor.

\subsection{Approximation of $r_{\mathrm{d}}(q)$}

During accretion in the stripping scenario, the distance $a$ between the components grows \citep{Blinnikov2022}, which causes the dimensionless radius of the massive component $r_1=R_1/a$ to decrease. At the same time the system becomes more asymmetric, and as can be seen from Fig.~\ref{Fig.rm_approx}, the minimum distance $\WI{r}{m}$ for which the stream approaches the massive component $M_1$ increases. So at some point a situation arises whereby the stream returns to strike itself at some radius $\WI{R}{cross}$, after reaching the minimum distance without having been stopped by the surface of the accretor, i.e. $r_1<\WI{r}{m}$. After self-intersection of the stream, an accretion disk starts to form \citep{LubowShu1975}.

\begin{figure}
	\includegraphics[width=\columnwidth]{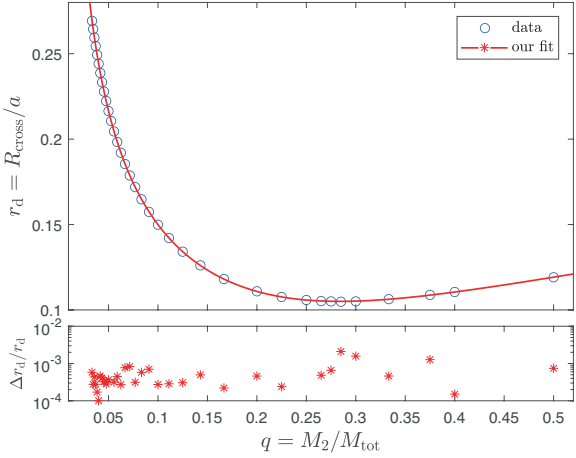}
	\caption{Upper panel: Comparison of numerical data with our approximation of $\WI{r}{d}(q)$. Lower panel: Deviations between our fit and computed data points ($\Delta \WI{r}{d} = |\mathrm{fit}-\mathrm{data}|$).} \label{Fig.r_disk}
\end{figure}

To qualitatively take into account the effect of orbital momentum transfer in formula $(\ref{J1_dot})$ during disk formation, one can use expression $(\ref{j_tot})$ for the specific angular momentum $\mathfrak{j}(q,\WI{r}{d }{=}R_{\mathrm{d}}/a)$. Here $R_{\mathrm{d}}$ is the characteristic radius of the accretion disk for which we chose the $\WI{R}{cross}$ mentioned above. In order to speed up our calculations, we also selected an approximation for the dimensionless radius of the stream's self-crossing point (see Fig.~\ref{Fig.r_disk}):
\begin{equation}
\WI{r}{d}=\frac{(a_1+a_2 q^{n_2})^{n_3}}{q^{n_1}}, \; 0.03\leqslant q \leqslant 0.5 \label{r_disk}
\end{equation}
where the coefficients are
\begin{center}
	\begin{tabular}{|c|c|}
		\hline
		$a_1$ & 2.7$\cdot 10^{-5}$\\ 
		\hline
		$a_2$ & 2.378$\cdot 10^{-3}$\\
		\hline
		$n_1$ & 5.4031$\cdot 10^{-1}$\\ 
		\hline
		$n_2$ & 3.514\\ 
		\hline
		$n_3$ & 2.994$\cdot 10^{-1}$\\ 
		\hline
	\end{tabular}
\end{center}
The accuracy of our approximation in the specified $q$ range is better than $0.5\%$, which is more than sufficient for estimating the effects of the orbital momentum transfer during disk formation.

\section{Discussion and Summary}
\label{4}

In this paper, we have studied the dynamics of direct impact accretion in a degenerate binary system at the final stages of its evolution. We have provided single-parameter approximations for the characteristic parameters of the system (the position of the L1 point, the distance of the minimum approach of the stream, etc.). The obtained expressions are significantly more accurate than known earlier approximations. For the first time, we have provided a two-parametric approximation for the specific angular momentum of the accreting matter transferring in the rotational momentum of the accretor.

\begin{figure}
	\includegraphics[width=\columnwidth]{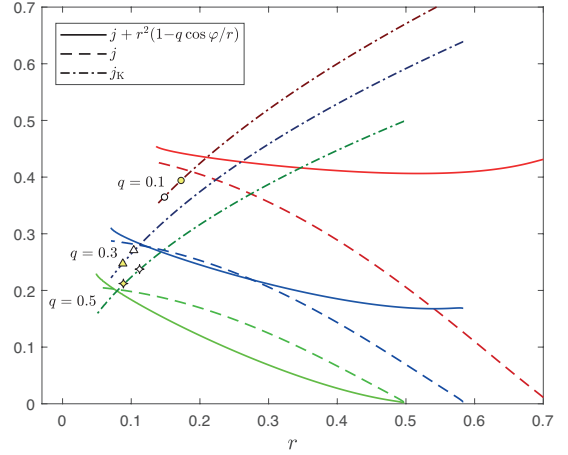} 
	\caption{The specific angular momentum of the accreting matter as a function of the dimensionless stopping radius $r$ for various $q=M_2/\WI{M}{tot}$. The solid line represents the behavior of moments in a non-rotating frame of reference, and the dashed line is in a rotating frame. The latter differs from the former by the kinematic correction associated with the transition from one frame of reference to another. The dashed-dotted line is the Keplerian angular momentum. See text for details.}  \label{Fig.j}
\end{figure}

In some works \citep[e.g.][]{PringleRees1972,LipunovPostnov1984,VerbuntRappaport1988} and also
\citet{Marsh2002,Marsh2004} where authors consider the case of direct impact accretion in WD-WD system, for simplicity, the specific angular momentum of the accreting matter $\mathfrak{j}$ in the formula (\ref{J1_dot}) is assumed to be equal to the Keplerian one:
\begin{equation}
\WI{j}{K}=\frac{\sqrt{G M_1 \WI{R}{1}}}{a^2 \Omega_{\mathrm{orb}}}=\sqrt{(1{-}q)\WI{r}{1}}.\label{J1_K}
\end{equation}
Fig.~\ref{Fig.j} shows the dependence of the specific momentum of matter from the stopping radius $r$ for three values of $q$. The solid line is our main approximation, the dashed line is the same without the kinematic correction, and the dash-dotted line is the Keplerian momentum. The white and yellow symbols on the Keplerian curves mark the value of the angular momentum at $r=\WI{r}{d}$ and $\WI{r}{h}$ \citep[see][]{VerbuntRappaport1988}, respectively. It can be seen from Fig.~\ref{Fig.j} that the Keplerian formula (\ref{J1_K}) gives not only a quantitative but also a qualitative discrepancy with formula (\ref{j_tot}) for $\mathfrak{j}$, a solution of the restricted 3-body problem, even for small $r$: while $\WI{j}{K}$ grows with the stopping radius $r$, $\mathfrak{j}$ is not monotonic. For large $q$ it decreases with increasing $r$, but starting from $q\approx 0.1$, $\mathfrak{j}$ practically does not depend on $r$.

The use of the obtained approximations allows us to refine the position of the mass boundary between the merging and stripping scenarios for WD-WD \citep{Marsh2002} and NS-NS \citep{ClarkEardley1977} systems, as well as the stripping time \citep[see also][]{Blinnikov2022}. It might also influence predictions of population synthesis for binary systems of degenerate stars with high mass asymmetry \citep[e.g.][]{Ferdman2020}.






\section*{Acknowledgements}
Authors are grateful to the RSF 21-12-00061 grant for support. We also thank the referee for useful comments.

\section*{Data Availability}

Data generated from computations are reported in the body of the paper. Additional data can be made available upon reasonable request.



\bibliographystyle{mnras}
\bibliography{direct_impact} 

\begin{thebibliography}{}
\makeatletter
\relax
\def\mn@urlcharsother{\let\do\@makeother \do\$\do\&\do\#\do\^\do\_\do\%\do\~}
\def\mn@doi{\begingroup\mn@urlcharsother \@ifnextchar [ {\mn@doi@}
  {\mn@doi@[]}}
\def\mn@doi@[#1]#2{\def\@tempa{#1}\ifx\@tempa\@empty \href
  {http://dx.doi.org/#2} {doi:#2}\else \href {http://dx.doi.org/#2} {#1}\fi
  \endgroup}
\def\mn@eprint#1#2{\mn@eprint@#1:#2::\@nil}
\def\mn@eprint@arXiv#1{\href {http://arxiv.org/abs/#1} {{\tt arXiv:#1}}}
\def\mn@eprint@dblp#1{\href {http://dblp.uni-trier.de/rec/bibtex/#1.xml}
  {dblp:#1}}
\def\mn@eprint@#1:#2:#3:#4\@nil{\def\@tempa {#1}\def\@tempb {#2}\def\@tempc
  {#3}\ifx \@tempc \@empty \let \@tempc \@tempb \let \@tempb \@tempa \fi \ifx
  \@tempb \@empty \def\@tempb {arXiv}\fi \@ifundefined
  {mn@eprint@\@tempb}{\@tempb:\@tempc}{\expandafter \expandafter \csname
  mn@eprint@\@tempb\endcsname \expandafter{\@tempc}}}

\bibitem[\protect\citeauthoryear{{Abbott} et~al.,}{{Abbott}
  et~al.}{2017a}]{Abbott2017a}
{Abbott} B.~P.,  et~al., 2017a, \mn@doi [\apjl] {10.3847/2041-8213/aa91c9},
  \href {https://ui.adsabs.harvard.edu/abs/2017ApJ...848L..12A} {848, L12}

\bibitem[\protect\citeauthoryear{{Abbott} et~al.,}{{Abbott}
  et~al.}{2017b}]{Abbott2017b}
{Abbott} B.~P.,  et~al., 2017b, \mn@doi [\apjl] {10.3847/2041-8213/aa920c},
  \href {https://ui.adsabs.harvard.edu/abs/2017ApJ...848L..13A} {848, L13}

\bibitem[\protect\citeauthoryear{{Blinnikov}, {Novikov}, {Perevodchikova}  \&
  {Polnarev}}{{Blinnikov} et~al.}{1984}]{Blinnikov1984}
{Blinnikov} S.~I.,  {Novikov} I.~D.,  {Perevodchikova} T.~V.,   {Polnarev}
  A.~G.,  1984, Soviet Astronomy Letters, \href
  {https://ui.adsabs.harvard.edu/abs/1984SvAL...10..177B} {10, 177}

\bibitem[\protect\citeauthoryear{{Blinnikov}, {Imshennik}, {Nadezhin},
  {Novikov}, {Perevodchikova}  \& {Polnarev}}{{Blinnikov}
  et~al.}{1990}]{Blinnikov1990}
{Blinnikov} S.~I.,  {Imshennik} V.~S.,  {Nadezhin} D.~K.,  {Novikov} I.~D.,
  {Perevodchikova} T.~V.,   {Polnarev} A.~G.,  1990, \sovast, \href
  {https://ui.adsabs.harvard.edu/abs/1990SvA....34..595B} {34, 595}

\bibitem[\protect\citeauthoryear{{Blinnikov}, {Nadyozhin}, {Kramarev}  \&
  {Yudin}}{{Blinnikov} et~al.}{2021}]{Blinnikov2021}
{Blinnikov} S.~I.,  {Nadyozhin} D.~K.,  {Kramarev} N.~I.,   {Yudin} A.~V.,
  2021, \mn@doi [Astronomy Reports] {10.1134/S1063772921050012}, \href
  {https://ui.adsabs.harvard.edu/abs/2021ARep...65..385B} {65, 385}

\bibitem[\protect\citeauthoryear{Blinnikov, Yudin, Kramarev  \&
  Potashov}{Blinnikov et~al.}{2022}]{Blinnikov2022}
Blinnikov S.,  Yudin A.,  Kramarev N.,   Potashov M.,  2022, \mn@doi
  [Particles] {10.3390/particles5020018}, 5, 198

\bibitem[\protect\citeauthoryear{{Clark} \& {Eardley}}{{Clark} \&
  {Eardley}}{1977}]{ClarkEardley1977}
{Clark} J.~P.~A.,  {Eardley} D.~M.,  1977, \mn@doi [\apj] {10.1086/155360},
  \href {https://ui.adsabs.harvard.edu/abs/1977ApJ...215..311C} {215, 311}

\bibitem[\protect\citeauthoryear{{Dan}, {Rosswog}, {Guillochon}  \&
  {Ramirez-Ruiz}}{{Dan} et~al.}{2011}]{Dan2011}
{Dan} M.,  {Rosswog} S.,  {Guillochon} J.,   {Ramirez-Ruiz} E.,  2011, \mn@doi
  [\apj] {10.1088/0004-637X/737/2/89}, \href
  {https://ui.adsabs.harvard.edu/abs/2011ApJ...737...89D} {737, 89}

\bibitem[\protect\citeauthoryear{{Dan}, {Rosswog}, {Guillochon}  \&
  {Ramirez-Ruiz}}{{Dan} et~al.}{2012}]{Dan2012}
{Dan} M.,  {Rosswog} S.,  {Guillochon} J.,   {Ramirez-Ruiz} E.,  2012, \mn@doi
  [\mnras] {10.1111/j.1365-2966.2012.20794.x}, \href
  {https://ui.adsabs.harvard.edu/abs/2012MNRAS.422.2417D} {422, 2417}

\bibitem[\protect\citeauthoryear{{Dan}, {Guillochon}, {Br{\"u}ggen},
  {Ramirez-Ruiz}  \& {Rosswog}}{{Dan} et~al.}{2015}]{Dan2015}
{Dan} M.,  {Guillochon} J.,  {Br{\"u}ggen} M.,  {Ramirez-Ruiz} E.,   {Rosswog}
  S.,  2015, \mn@doi [\mnras] {10.1093/mnras/stv2289}, \href
  {https://ui.adsabs.harvard.edu/abs/2015MNRAS.454.4411D} {454, 4411}

\bibitem[\protect\citeauthoryear{{Eichler}, {Livio}, {Piran}  \&
  {Schramm}}{{Eichler} et~al.}{1989}]{Eichler1989}
{Eichler} D.,  {Livio} M.,  {Piran} T.,   {Schramm} D.~N.,  1989, \mn@doi
  [\nat] {10.1038/340126a0}, \href
  {https://ui.adsabs.harvard.edu/abs/1989Natur.340..126E} {340, 126}

\bibitem[\protect\citeauthoryear{{Ferdman} et~al.,}{{Ferdman}
  et~al.}{2020}]{Ferdman2020}
{Ferdman} R.~D.,  et~al., 2020, \mn@doi [\nat] {10.1038/s41586-020-2439-x},
  \href {https://ui.adsabs.harvard.edu/abs/2020Natur.583..211F} {583, 211}

\bibitem[\protect\citeauthoryear{{Flannery}}{{Flannery}}{1975}]{Flannery1975}
{Flannery} B.~P.,  1975, \mn@doi [\mnras] {10.1093/mnras/170.2.325}, \href
  {https://ui.adsabs.harvard.edu/abs/1975MNRAS.170..325F} {170, 325}

\bibitem[\protect\citeauthoryear{{Gokhale}, {Peng}  \& {Frank}}{{Gokhale}
  et~al.}{2007}]{2007ApJ.Gokhale}
{Gokhale} V.,  {Peng} X.~M.,   {Frank} J.,  2007, \mn@doi [\apj]
  {10.1086/510119}, \href
  {https://ui.adsabs.harvard.edu/abs/2007ApJ...655.1010G} {655, 1010}

\bibitem[\protect\citeauthoryear{{Guillochon}, {Dan}, {Ramirez-Ruiz}  \&
  {Rosswog}}{{Guillochon} et~al.}{2010}]{Guillochon2010}
{Guillochon} J.,  {Dan} M.,  {Ramirez-Ruiz} E.,   {Rosswog} S.,  2010, \mn@doi
  [\apjl] {10.1088/2041-8205/709/1/L64}, \href
  {https://ui.adsabs.harvard.edu/abs/2010ApJ...709L..64G} {709, L64}

\bibitem[\protect\citeauthoryear{{Gurfil} \& {Seidelmann}}{{Gurfil} \&
  {Seidelmann}}{2016}]{Gurfil_book}
{Gurfil} P.,  {Seidelmann} P.~K.,  2016, Celestial Mechanics and Astrodynamics:
  Theory and Practice.
Springer Berlin, Heidelberg, \url
  {https://link.springer.com/book/10.1007/978-3-662-50370-6#bibliographic-information}

\bibitem[\protect\citeauthoryear{Haensel, Potekhin  \& Yakovlev}{Haensel
  et~al.}{2007}]{Haensel_book}
Haensel P.,  Potekhin A.,   Yakovlev D.,  2007, Neutron Stars 1: Equation of
  State and Structure.
Astrophysics and Space Science Library, Springer New York, \url
  {https://books.google.ru/books?id=fgj\_TZ06niYC}

\bibitem[\protect\citeauthoryear{{Iben} \& {Tutukov}}{{Iben} \&
  {Tutukov}}{1984}]{Iben1984ApJS}
{Iben} I. J.,  {Tutukov} A.~V.,  1984, \mn@doi [\apjs] {10.1086/190932}, \href
  {https://ui.adsabs.harvard.edu/abs/1984ApJS...54..335I} {54, 335}

\bibitem[\protect\citeauthoryear{{Kaplan}, {Nichols}  \& {Thorne}}{{Kaplan}
  et~al.}{2009}]{Kaplan2009}
{Kaplan} J.~D.,  {Nichols} D.~A.,   {Thorne} K.~S.,  2009, \mn@doi [\prd]
  {10.1103/PhysRevD.80.124014}, \href
  {https://ui.adsabs.harvard.edu/abs/2009PhRvD..80l4014K} {80, 124014}

\bibitem[\protect\citeauthoryear{{Kashyap}, {Haque}, {Lor{\'e}n-Aguilar},
  {Garc{\'\i}a-Berro}  \& {Fisher}}{{Kashyap} et~al.}{2018}]{Kashyap2018}
{Kashyap} R.,  {Haque} T.,  {Lor{\'e}n-Aguilar} P.,  {Garc{\'\i}a-Berro} E.,
  {Fisher} R.,  2018, \mn@doi [\apj] {10.3847/1538-4357/aaedb7}, \href
  {https://ui.adsabs.harvard.edu/abs/2018ApJ...869..140K} {869, 140}

\bibitem[\protect\citeauthoryear{{Koester} \& {Chanmugam}}{{Koester} \&
  {Chanmugam}}{1990}]{Koester1990}
{Koester} D.,  {Chanmugam} G.,  1990, \mn@doi [Reports on Progress in Physics]
  {10.1088/0034-4885/53/7/001}, \href
  {https://ui.adsabs.harvard.edu/abs/1990RPPh...53..837K} {53, 837}

\bibitem[\protect\citeauthoryear{{Kopal}}{{Kopal}}{1959}]{Kopal_book}
{Kopal} Z.,  1959, Close binary systems.
New York: Wiley, \url
  {https://archive.org/details/closebinarysyste00kopa/mode/2up}

\bibitem[\protect\citeauthoryear{{Kowalska}, {Bulik}, {Belczynski}, {Dominik}
  \& {Gondek-Rosinska}}{{Kowalska} et~al.}{2011}]{Kowalska2011}
{Kowalska} I.,  {Bulik} T.,  {Belczynski} K.,  {Dominik} M.,
  {Gondek-Rosinska} D.,  2011, \mn@doi [\aap] {10.1051/0004-6361/201015777},
  \href {https://ui.adsabs.harvard.edu/abs/2011A&A...527A..70K} {527, A70}

\bibitem[\protect\citeauthoryear{{Lattimer} \& {Schramm}}{{Lattimer} \&
  {Schramm}}{1974}]{Lattimer1974ApJ}
{Lattimer} J.~M.,  {Schramm} D.~N.,  1974, \mn@doi [\apjl] {10.1086/181612},
  \href {https://ui.adsabs.harvard.edu/abs/1974ApJ...192L.145L} {192, L145}

\bibitem[\protect\citeauthoryear{{Lenon}, {Nitz}  \& {Brown}}{{Lenon}
  et~al.}{2020}]{Lenon2020}
{Lenon} A.~K.,  {Nitz} A.~H.,   {Brown} D.~A.,  2020, \mn@doi [\mnras]
  {10.1093/mnras/staa2120}, \href
  {https://ui.adsabs.harvard.edu/abs/2020MNRAS.497.1966L} {497, 1966}

\bibitem[\protect\citeauthoryear{{Lipunov} \& {Postnov}}{{Lipunov} \&
  {Postnov}}{1984}]{LipunovPostnov1984}
{Lipunov} V.~M.,  {Postnov} K.~A.,  1984, \mn@doi [\apss] {10.1007/BF00653919},
  \href {https://ui.adsabs.harvard.edu/abs/1984Ap&SS.106..103L} {106, 103}

\bibitem[\protect\citeauthoryear{{Lubow} \& {Shu}}{{Lubow} \&
  {Shu}}{1975}]{LubowShu1975}
{Lubow} S.~H.,  {Shu} F.~H.,  1975, \mn@doi [\apj] {10.1086/153614}, \href
  {https://ui.adsabs.harvard.edu/abs/1975ApJ...198..383L} {198, 383}

\bibitem[\protect\citeauthoryear{{Marsh} \& {Steeghs}}{{Marsh} \&
  {Steeghs}}{2002}]{Marsh2002}
{Marsh} T.~R.,  {Steeghs} D.,  2002, \mn@doi [\mnras]
  {10.1046/j.1365-8711.2002.05346.x}, \href
  {https://ui.adsabs.harvard.edu/abs/2002MNRAS.331L...7M} {331, L7}

\bibitem[\protect\citeauthoryear{{Marsh}, {Nelemans}  \& {Steeghs}}{{Marsh}
  et~al.}{2004}]{Marsh2004}
{Marsh} T.~R.,  {Nelemans} G.,   {Steeghs} D.,  2004, \mn@doi [\mnras]
  {10.1111/j.1365-2966.2004.07564.x}, \href
  {https://ui.adsabs.harvard.edu/abs/2004MNRAS.350..113M} {350, 113}

\bibitem[\protect\citeauthoryear{{Murray} \& {Dermott}}{{Murray} \&
  {Dermott}}{1999}]{1999MurrayDermott.book}
{Murray} C.~D.,  {Dermott} S.~F.,  1999, {Solar system dynamics}

\bibitem[\protect\citeauthoryear{{Nelemans}, {Yungelson}, {Portegies Zwart}  \&
  {Verbunt}}{{Nelemans} et~al.}{2001a}]{Nelemans2001a}
{Nelemans} G.,  {Yungelson} L.~R.,  {Portegies Zwart} S.~F.,   {Verbunt} F.,
  2001a, \mn@doi [\aap] {10.1051/0004-6361:20000147}, \href
  {https://ui.adsabs.harvard.edu/abs/2001A&A...365..491N} {365, 491}

\bibitem[\protect\citeauthoryear{{Nelemans}, {Portegies Zwart}, {Verbunt}  \&
  {Yungelson}}{{Nelemans} et~al.}{2001b}]{Nelemans2001b}
{Nelemans} G.,  {Portegies Zwart} S.~F.,  {Verbunt} F.,   {Yungelson} L.~R.,
  2001b, \mn@doi [\aap] {10.1051/0004-6361:20010049}, \href
  {https://ui.adsabs.harvard.edu/abs/2001A&A...368..939N} {368, 939}

\bibitem[\protect\citeauthoryear{{Plavec} \& {Kratochvil}}{{Plavec} \&
  {Kratochvil}}{1964}]{Plavec1964}
{Plavec} M.,  {Kratochvil} P.,  1964, Bulletin of the Astronomical Institutes
  of Czechoslovakia, \href
  {https://ui.adsabs.harvard.edu/abs/1964BAICz..15..165P} {15, 165}

\bibitem[\protect\citeauthoryear{{Pringle} \& {Rees}}{{Pringle} \&
  {Rees}}{1972}]{PringleRees1972}
{Pringle} J.~E.,  {Rees} M.~J.,  1972, \aap, \href
  {https://ui.adsabs.harvard.edu/abs/1972A&A....21....1P} {21, 1}

\bibitem[\protect\citeauthoryear{{Sepinsky} \& {Kalogera}}{{Sepinsky} \&
  {Kalogera}}{2014}]{Sepinsky2014ApJ}
{Sepinsky} J.~F.,  {Kalogera} V.,  2014, \mn@doi [\apj]
  {10.1088/0004-637X/785/2/157}, \href
  {https://ui.adsabs.harvard.edu/abs/2014ApJ...785..157S} {785, 157}

\bibitem[\protect\citeauthoryear{{Sepinsky}, {Willems}, {Kalogera}  \&
  {Rasio}}{{Sepinsky} et~al.}{2010}]{Sepinsky2010ApJ}
{Sepinsky} J.~F.,  {Willems} B.,  {Kalogera} V.,   {Rasio} F.~A.,  2010,
  \mn@doi [\apj] {10.1088/0004-637X/724/1/546}, \href
  {https://ui.adsabs.harvard.edu/abs/2010ApJ...724..546S} {724, 546}

\bibitem[\protect\citeauthoryear{Shapiro \& Teukolsky}{Shapiro \&
  Teukolsky}{2008}]{ShapiroTeukolsky_book}
Shapiro S.,  Teukolsky S.,  2008, Black Holes, White Dwarfs, and Neutron Stars:
  The Physics of Compact Objects.
Wiley, \url {https://books.google.ru/books?id=d1CRQIcP1zoC}

\bibitem[\protect\citeauthoryear{{Silber}}{{Silber}}{1992}]{Silber1992}
{Silber} A.~D.,  1992, PhD thesis, Massachusetts Institute of Technology

\bibitem[\protect\citeauthoryear{Tassoul}{Tassoul}{1978}]{Tassoul1978}
Tassoul J.-L.,  1978, Theory of Rotating Stars.
Princeton University Press, \url {http://www.jstor.org/stable/j.ctt13x0sgx}

\bibitem[\protect\citeauthoryear{{Verbunt} \& {Rappaport}}{{Verbunt} \&
  {Rappaport}}{1988}]{VerbuntRappaport1988}
{Verbunt} F.,  {Rappaport} S.,  1988, \mn@doi [\apj] {10.1086/166645}, \href
  {https://ui.adsabs.harvard.edu/abs/1988ApJ...332..193V} {332, 193}

\bibitem[\protect\citeauthoryear{Warner}{Warner}{1995}]{Warner_book}
Warner B.,  1995, Cataclysmic Variable Stars.
Cambridge Astrophysics, Cambridge University Press,
  \mn@doi{10.1017/CBO9780511586491}

\bibitem[\protect\citeauthoryear{{Warner} \& {Peters}}{{Warner} \&
  {Peters}}{1972}]{WarnerPeters1972}
{Warner} B.,  {Peters} W.~L.,  1972, \mn@doi [\mnras] {10.1093/mnras/160.1.15},
  \href {https://ui.adsabs.harvard.edu/abs/1972MNRAS.160...15W} {160, 15}

\bibitem[\protect\citeauthoryear{{Webbink}}{{Webbink}}{1984}]{Webbink1984ApJ}
{Webbink} R.~F.,  1984, \mn@doi [\apj] {10.1086/161701}, \href
  {https://ui.adsabs.harvard.edu/abs/1984ApJ...277..355W} {277, 355}

\bibitem[\protect\citeauthoryear{{Yudin}, {Razinkova}  \& {Blinnikov}}{{Yudin}
  et~al.}{2020}]{Yudin2020}
{Yudin} A.~V.,  {Razinkova} T.~L.,   {Blinnikov} S.~I.,  2020, \mn@doi
  [Astronomy Letters] {10.1134/S1063773719120077}, \href
  {https://ui.adsabs.harvard.edu/abs/2020AstL...45..847Y} {45, 847}

\makeatother
\end{thebibliography}




\appendix

\section{The effect of non-sphericity}
\label{Appendix0}

Let us estimate the effect of non-sphericity of the accretor and its influence on the external gravitational potential. In the case of axisymmetric barotropes, the equilibrium equation can be written as \citep[e.g.][]{Tassoul1978}
\begin{equation}
H+U=\int\!\!\omega^2\xi d\xi,
\end{equation}
where $H=\int \!dP/\rho$ is enthalpy of matter as function of its pressure $P$ and density $\rho$, $U$ is the gravitational potential, $\omega$ is the angular velocity of rotation, and $\xi$ is the cylindrical radius. We also assume that the rotation is rigid with $\omega=\mathrm{const}$. On the surface of the accretor $H=0$ so we can obtain
\begin{equation}
U(\re)=U(\rp)+\frac{\omega^2\re^2}{2},\label{phi_e_p}
\end{equation}
where $\WI{R}{p,e}$ are the polar and equatorial radii of the rotating star, respectively. In the case of the limiting Keplerian rotation, the following force balance equation also holds at the equator:
\begin{equation}
\nabla U(\re)=\omega^2\re.\label{critical_rotation}
\end{equation}
The gravitational potential as well as its derivatives are continuous at the boundary of the star. For the external potential we use the well-known spherical harmonics expansion and leave only the first two terms in this expansion:
\begin{equation}
U=-\frac{GM}{R}+\frac{B}{R^3}P_2(\mu),\label{phi_out}
\end{equation}
where $\mu$ is the cosine of the angle between the radius vector $R$ and the spin axis, and $P_2=(3\mu^2{-}1)/2$ is the Legendre polynomial. Substituting (\ref{phi_out}) into the relations (\ref{phi_e_p}) and (\ref{critical_rotation}), we obtain the expression for the external potential in the form:
\begin{equation}
U=-\frac{GM}{R}\bigg[1+P_2(\mu)\bigg(\frac{\re}{R}\bigg)^2\frac{\theta^2(2{-}3\theta)}{4{+}5\theta^3}\bigg],\label{phi_out_sol}
\end{equation}
where we introduce the notation $\theta\equiv\rp/\re$. The rapidly rotating polytropes with indices $n\sim (1\div 3)$ have $\theta\sim(\frac{1}{2}\div\frac{2}{3})$, the case $\theta\ll 1$ can be also provided by a differential rotation \citep{Tassoul1978}. As can be seen from (\ref{phi_out_sol}), the correction to the spherically symmetric potential is small at $R\gg\re$ and does not influence the parameters of the binary system. However, the dynamics of a stream of matter near the surface of the strongly spun up accretor may be changed.

\section{GR effects}
\label{Appendix}

In deriving the equations (\ref{energy4}-\ref{moment4}), we do not take into account the GR effects. They are mainly related to frame-dragging, because the massive component can be strongly spun up during accretion \citep{Blinnikov2022}. Let us write an expression for the gravitational potential created by a point mass $M_1$ with spin $J_1$ at a distance $R$ in the post-Newtonian approximation \citep{Kaplan2009}:
\begin{equation}
\WI{U}{PN}=\WI{U}{N}\left(1+\frac{3}{2}\frac{V^2}{c^2}-\frac{G M_2}{c^2 a}\right)+2\frac{G V J_1}{c^2 R^2},\label{U_Kaplan}
\end{equation}
where $\WI{U}{N}=G M_1/R$ is the Newtonian potential, $V$ is the modulus of the particle velocity of the accreting matter, and $a$ is the distance between the stars. Calculations show that $V/c \lesssim 0.1$ near the surface of the accretor ($R_1 \ll a$), so the main addition to the Newtonian potential should come from the last term, which is just related to frame-dragging.

Let us obtain an upper bound for this effect by considering the case of the limiting Keplerian rotation of a massive NS. Its rotational momentum in this case is
\begin{equation}
J_1 = I_1 \Omega_K,
\end{equation}
where $I_1 = \beta M_1 R_1^2$ with numerical parameters $\beta=O(1)$ \citep[e.g.][]{Haensel_book} and $\Omega_K^2 \leq \pi G \bar{\rho }$ ($\bar{\rho}\sim M_1/R_1^3$ is average density of NS) \citep[e.g.][]{Tassoul1978}. Setting $M_1 = 2 M_{\odot}$ and $R=R_1=10$ km we get
\begin{equation}
\frac{|\WI{U}{PN}{-}\WI{U}{N}|}{\WI{U}{N}}\sim \sqrt{\frac{G M_1}{c^2 R_1}} \frac{V}{c}\lesssim \frac{V}{c}.
\end{equation}
Considering the case of an extremely rotating massive NS, we obtain an upper bound for the contribution of GR effects near the NS surface of about 10\%.


\bsp	
\label{lastpage}
\end{document}